\documentclass[%
10pt,
a4paper,
twocolumn,
aps,prb,
superscriptaddress,
showpacs
]{revtex4-2}

\usepackage[utf8]{inputenc}
\usepackage[english]{babel}
\usepackage{dirtytalk}
\usepackage{amsfonts}
\usepackage{amsmath}
\usepackage{amssymb}
\usepackage{bbold}
\usepackage[dvipsnames,svgnames,table]{xcolor}
\usepackage{graphicx}
\usepackage{fancyhdr}
\usepackage{float}
\usepackage{braket}
\usepackage{float}
\usepackage[caption=false]{subfig}

\usepackage{blindtext}

\usepackage{silence}
\usepackage{tikz}

\usepackage{hyperref}
\hypersetup{
    pdfstartview={FitH},
    colorlinks=true,    
    linkcolor=NavyBlue, 
    citecolor=Maroon,   
    filecolor=NavyBlue, 
    urlcolor=NavyBlue   
}

\makeatletter
\def\footnoterule{\kern -10pt
    \hrule \@width 100pt \kern 10pt} 
\makeatother

\makeatletter
    \def\bbl@set@language#1{%
      \edef\languagename{%
        \ifnum\escapechar=\expandafter`\string#1\@empty
        \else\string#1\@empty\fi}%
      \@ifundefined{babel@language@alias@\languagename}{}{%
        \edef\languagename{\@nameuse{babel@language@alias@\languagename}}%
      }%
      \select@language{\languagename}%
      \expandafter\ifx\csname date\languagename\endcsname\relax\else
        \if@filesw
          \protected@write\@auxout{}{\string\select@language{\languagename}}%
          \bbl@for\bbl@tempa\BabelContentsFiles{%
            \addtocontents{\bbl@tempa}{\xstring\select@language{\languagename}}}%
          \bbl@usehooks{write}{}%
        \fi
      \fi}
    \newcommand{\DeclareLanguageAlias}[2]{%
      \global\@namedef{babel@language@alias@#1}{#2}%
    }
\makeatother
\DeclareLanguageAlias{en}{english}

\def \jgu {Insitut für Physik, Johannes Gutenberg-Universität Mainz, D-55099, Germany}

\bibpunct{[}{]}{,}{n}{}{}  

\begin{document}

    \title{Strain controlled \textit{g}- to \textit{d}-wave transition in altermagnetic CrSb}
        
    \author{Bennet Karetta}
        \thanks{These two authors contributed equally to this work.}
        \affiliation{\jgu}

    \author{Xanthe H. Verbeek}
        \thanks{These two authors contributed equally to this work.}
        \affiliation{\jgu}

    \author{Rodrigo Jaeschke-Ubiergo}
        \affiliation{\jgu}

    \author{Libor Šmejkal}
        \affiliation{Max-Planck-Institut für Physik komplexer Systeme, Nöthnitzer Str. 38, 01187 Dresden, Germany}
        \affiliation{Max-Planck-Institut für Chemische Physik fester Stoffe, Nöthnitzer Str. 40, 01187 Dresden, Germany}
        \affiliation{Inst. of Physics Academy of Sciences of the Czech Republic, Cukrovarnick\'{a} 10,  Praha 6, Czech Republic}

    \author{Jairo Sinova}
    \email{sinova@uni-mainz.de}
    \affiliation{\jgu}
    \affiliation{Department of Physics, Texas A\&M University, College Station, Texas 77843-4242, USA}

\begin{abstract}
The possibility of a strain-induced transformation from $g$-wave to $d$-wave altermagnetism was recently recently proposed for MnTe using a $k\cdot p$ perturbative model. In this work, we demonstrate such a transition in CrSb for a wider array of strains, using a combination of a minimal model and first-principles calculations. Starting from a symmetry perspective, we analyze the spin elastoconductivity tensor, and determine the strain types which allow for a change in the altermagnetic symmetry. We obtain three strain directions, which allow for a $d$-wave type splitting, and one in which a net magnetic moment emerges. Using first-principles calculations in the absence of spin-orbit coupling (SOC), we confirm these symmetry predictions. Furthermore, these results do not alter qualitatively in the presence of SOC. Finally, we reveal that the resulting spin currents give rise to a spin-splitter effect of up to 5\% under realistic strains of 1\%, confirming strain as a powerful tool for tuning altermagnetic properties.
\end{abstract}

\maketitle

\section{Introduction}
\label{sec::introduction}

Altermagnets form a newly recognized class of collinear magnets, showing spin ordering in both direct and momentum space and exhibiting $d$-, $g$- or $i$-wave symmetry \cite{Smejkal2021a}. Similar to antiferromagnets, altermagnets are magnetically compensated, but, like ferromagnets, they break both time-reversal symmetry (TRS) and the product of space inversion and TRS, such that Kramers spin degeneracy is lifted. Altermagnets are characterized a by spin order with $d$-, $g$- or $i$-wave symmetry of both: (i) the spin splitting of the non-relativistic electronic band structure in momentum space \cite{Smejkal2021a,Smejkal2022a}; (ii) the ferroic order of the atomic spin density in direct space \cite{Jungwirth2024, Jaeschke2025, Bhowal2024, Fernandes2023, Verbeek2024}.

Altermagnetism was predicted through a detailed classification and delineation of collinear magnetic phases based on spin symmetries \cite{Smejkal2021a}. These symmetries involve pairs of distinct operations that act on both the lattice and spin degrees of freedom. As a result, altermagnetism leads to an unconventional, complex spin density \cite{Smejkal2020} and exchange fields \cite{Smejkal2023, Hoyer2024}, which break the lattice symmetry, in a manner similar to the behavior observed in unconventional superfluid states \cite{Smejkal2022a, Jungwirth2024b}. These spin symmetries consider the non-relativistic physics, rather than the magnetic space group symmetries, which take also relativistic physics into account. The spin symmetry analysis, which strictly speaking holds in the non-relativistic limit, is well suited to describe the exchange-dominated physics, even in systems with high spin-orbit coupling (SOC) \cite{Pari2024}. 

Recent experimental observations have confirmed the presence of altermagnetic splitting with $g$-wave symmetry in both MnTe and CrSb, using photoemission spectroscopy of the electronic band structure \cite{Krempasky2023,Lee2024,Osumi2024,Reimers2024,Yang2024,Ding2024,Zeng2024,Li2024,Hariki2023, Amin2024}. 
On the other hand, the search for metallic $d$-wave altermagnetic candidates has been more elusive. Despite the experimental evidence of anomalous Hall and spin currents in RuO$_2$, there is still an open debate about its magnetic order \cite{Berlijn2017a,Zhu2018,Lovesey2022,Occhialini2021,Feng2022,Bose2022,Bai2022,Karube2022,Lovesey2023c,Liu2023,Fedchenko2024,Smolyanyuk2023,Lin2024,Kessler2024,Li2024a,Wenzel2025,Jeong2024,Hiraishi2024}.
More recently, nuclear magnetic resonance and neutron diffraction measurements, guided by ab initio techniques, have enabled the identification of KV$_2$Se$_2$O, Rb$_2$V$_2$Te$_2$O and CoNb$_4$Se$_8$ as metallic $d$-wave altermagnets \cite{Jiang2025,Zhang2025a, Regmi2024}.

The identification of the altermagnetic spin symmetry class has also provided a unified explanation for previous reports of unconventional TRS breaking in electronic structures and related phenomena, including the anomalous Hall effect \cite{Smejkal2020, Mazin2021}, spin currents \cite{Ahn2019, Naka2019, Gonzalez-Hernandez2021,Smejkal2022GMR}, and magneto-optical effects \cite{Mazin2021, Samanta2020}. The unique electronic structure of altermagnets features TRS breaking spin-split bands across the entire Brillouin Zone, except along nodal surfaces corresponding to d-wave (2), g-wave (4), or i-wave (6) symmetries \cite{Smejkal2020, Mazin2021}. This leads to a variety of spintronic effects \cite{Smejkal2020, Mazin2021, Naka2019, Smejkal2022GMR, Feng2022, Reichlova2024, Bai2024, Shao2021}. One of these is the spin-splitter effect, in which an electric bias induces a pure transverse spin current, happens naturally in $d$-wave altermagnets, but it is forbidden in the $g$, and $i$- wave counterparts. This effect is especially significant for applications in spintronics, as it allows for the generation of pure spin currents. 

Considering the important role symmetry plays in both the separation of altermagnets from other types of collinear magnets, as well as in determining the type of altermagnetism, it seems attractive to manipulate the symmetry to induce a change in the spin splitting and the associated effects. 
For example, recently spin currents were predicted in strained doped MnTe \cite{Belashchenko2024}. Here, the combination of the in-plane orientation of the Néel vector and the strain reduced the symmetries of the original $g$-wave spin splitting, making such a response possible. 

In this article, we focus on CrSb, a $g$-wave altermagnet with an out-of-plane N\'eel vector, and explore a range of strains for which a symmetry breaking can occur. CrSb is a metallic $g$-wave altermagnet, with a high ordering temperature $T_{N} =700$ K \cite{Takei1963}. The crystal structure of CrSb is hexagonal, with space group $P6_{3}/mmc$ (No. 194), and the magnetic easy axis is parallel to the crystallographic c-axis \cite{Snow1952}. This makes it different from MnTe, whose in-plane anisotropy allows for a spontaneous AHE and other axial magnetoelectrical responses allowed by this magnetic symmetry. 
In addition, the spin-splitting in CrSb is large, reaching 0.6 eV near the Fermi energy \cite{Reimers2024}, but as it is a $g$-wave altermagnet, there is no spin-splitter effect in the unstrained material. 

In CrSb, the magnetic compensated order has ferromagnetic (001) planes of Cr atoms with alternating spin polarization along the c-axis. This pattern of spin splitting is described by the non-trivial spin point group 
\begin{align}
    ^{2}6/^{2}m^{2}m^{1}m = [E||\bar{3}m] + [C_2||C_{6z}\bar{3}m] .
\end{align}
In the notation $[\,\cdot\,||\,\cdot\,]$ the operations on the left of vertical bars ($E$ and $C_2$) act only in spin space. The operations on the right of the vertical bar act only in real space. In this case, all the crystallographic symmetries contained in the subgroup $\bar{3}m$ connect each spin sublattice with itself. On the other hand, all symmetries contained in coset $C_{6z}\bar{3}m$ map the two sublattices into each other. Thus, these are only a symmetry of the full system when combined with a two-fold spin rotation. Out of these symmetries, we will specifically focus on $[C_{2}||C_{6z}]$, $[C_2||m_z]$ and $[C_2||m_y]$, and when these are preserved or broken.

In this work, we explore a set of shear strains for which we predict the emergence of the spin-splitter effect in the $g$-wave altermagnet CrSb, showing a strain-induced spin-splitter effect. Starting with a symmetry analysis, we show that the application of shear strain will reduce the crystal symmetry, resulting in different point groups, depending on the strain direction. Specifically, the symmetry of the new spin point group determines which, if any, of the components of the spin conductivity tensor are allowed to be non-zero, and thus if a spin-splitter effect is allowed. Based on this symmetry analysis, the strain directions of interest are selected. 

In the next section, Sec.~\ref{sec::minimal_model}, we explore the effect of these strains using a minimal model, capturing the basic altermagnetic features of CrSb. In Sec.~\ref{sec::CrSb} we investigate these features in a full \textit{ab initio} description of CrSb, and compare results with and without spin-orbit coupling. In Sec.~\ref{sec::spin-elastoconductivity} we discuss the spin elastoconductivity and 
in Sec.~\ref{sec::conclusion} we present our conclusions. 

\section{Minimal Model}
\label{sec::minimal_model}

To illustrate the symmetry breaking from $g$-wave to $d$-wave altermagnetism, we start by considering a minimal two-band model satisfying the symmetries of the spin point group (SPG) $^26/^2m^2m^1m$. As we will see later, the crystal structure of CrSb undergoes the same transitions that are captured by this model. We study the two-band model near the $\Gamma $-point, where the low-energy effective Hamiltonian in momentum space takes the form:

\begin{align}
    \mathcal{H}_0(\mathbf{k}) = t_0 |\mathbf{k}|^2 \tau_0 + t k_y k_z (3k_x^2 - k_y^2) \tau_z.
\end{align}
Here, $\tau_0$ denotes the identity matrix and $\tau_z$ is the $z$-component of the Pauli matrices, both acting in spin space. $t_0$ describes the strength of the non-magnetic influences on the band structure and competes with $t$, the strength of the altermagnetic spin splitting. As this system has $g$-wave altermagnetic symmetry, the generation of spin currents from the spin-splitter effect is not allowed. In other words, the spin conductivity is zero.

Analogously to the charge conductivity, we can describe the \emph{spin} current propagation through a material with the spin conductivity tensor. Specifically, in the collinear non-relativistic regime, the spin is a good quantum number, and the two spin channels (up/$\uparrow$ and down/$\downarrow$) along the spin quantization axis are decoupled, so the charge conductivity $\hat{\sigma}^{c}$ can be separated in $\hat{\sigma}^{c,\uparrow}$ and $\hat{\sigma}^{c,\downarrow}$, with $\hat{\sigma}^c = \hat{\sigma}^{c,\uparrow}+\hat{\sigma}^{c,\downarrow}$. On the other hand, subtracting both conductivities will give the spin conductivity, and we define the spin conductivity tensor $\hat{\sigma}^s$ as:
\begin{align}
   \hat{\sigma}^s =  \frac{\hbar}{2e} (\hat{\sigma}^{c,\uparrow}-\hat{\sigma}^{c,\downarrow}),
\end{align}
where $\hbar$ is the reduced Planck constant and $e$ is the elementary charge. 

To investigate symmetry-breaking transitions from the $g$-wave altermagnetic phase to either a $d$-wave altermagnetic state or an uncompensated state, we explore the conditions under which spin currents can be induced via strain. To this end we study the spin elastoconductivity tensor $\hat{\Gamma}$, a four-rank tensor describing the change in the spin conductivity tensor $\hat{\sigma}^s$ from its unstrained value $\hat{\sigma}^s_0$ for small strains $\hat{u}$. We can write the relation between $\hat{\sigma}^s$, $\hat{\sigma}^s$ and $\hat{\Gamma}$ as:

\begin{align}
    \hat{\sigma}^s = \hat{\sigma}^s_0 + \left( \sum_{k,l = x,y,z} \Gamma_{ij,kl} u_{kl} \right)_{(i,j = x,y,z)} .
\end{align}

{ A detailed description of how to obtain the $\hat{\Gamma}$-tensor and a table with the spin elastoconductivity tensor for all altermagnetic spin-Laue-groups can be found in the Supplemental Material}.

Under the symmetry constraints of the spin point group $^26/^2m^2m^1m$, the tensor $\hat{\Gamma}$ reveals that four specific shear strain configurations of the strain tensor $\hat{u}$ can induce non-zero spin currents. These include the anisotropic diagonal shear $\hat{u}_{xx-yy}$, defined by $u = u_{xx} = -u_{yy}$ with all other components vanishing, as well as the off-diagonal shear strains $\hat{u}_{ij}$ for all $i \neq j$, with $u = u_{ij} = u_{ji}$ and all other components zero. To investigate our model under strain we perform the following transformation of the $\mathbf{k}$-vector:
$ \mathbf{k} \rightarrow ( 1 + \hat{u})\cdot \mathbf{k} $, which leads to a modified Hamiltonian
\begin{align}
    \mathcal{H} (\mathbf{k}, u ) = \mathcal{H}_0(\mathbf{k}) + \mathcal{H}_{\hat{u}} (\mathbf{k}) u + \mathcal{O}\left(u^2\right),
\end{align}
where $\mathcal{H}_{\hat{u}}(\mathbf{k})$ captures the first-order corrections due to strain which we will focus on here. 

Table \ref{tab::toy_model_table} presents those terms in the Hamiltonian $\mathcal{H}$ that are linear in strain along with their corresponding SPG symmetries for each of the four relevant strain configurations. As can be seen from the table, the new SPGs have significantly reduced symmetries, no longer enough to support the $g$-wave spin splitting pattern, and can be classified as either altermagnetic or ferromagnetic. This shows the application of strain can induce a symmetry-lowering transition from the original $g$-wave altermagnetic state to either a $d$-wave altermagnetic configuration or an uncompensated phase, showing that the symmetry reduction under the application of shear strain can also lead to the emergence of a net magnetic moment. 
\begin{table}[t]
    \centering
    \small{
    \begin{tabular}{ccccccc}
        \hline
        \\
        Strain type & \hspace{0.1cm} & $\mathcal{H}_{\hat{u}}(\mathbf{k})$ & \hspace{0.1cm} & SPG & \hspace{0.1cm}  & class  \\
        \\
        \hline \hline
        \\
        $\hat{u}_{xx-yy}$ & \hspace{0.1cm} & $k_y k_z (k_x^2 + k_y^2)$                 & \hspace{0.1cm} & $^2m^2m^1m$ & \hspace{0.1cm} & AM \\
        \\
        \hline
        \\
        $\hat{u}_{xy}$    & \hspace{0.1cm} & $k_x k_z (k_x^2 + k_y^2)$                 & \hspace{0.1cm} & $^22/^2m$   & \hspace{0.1cm} & AM \\
        \\
        \hline
        \\
        $\hat{u}_{zx}$    & \hspace{0.1cm} & $k_x k_z (3k_x^2 - k_y^2 + 6k_z^2)$       & \hspace{0.1cm} & $^22/^2m$   & \hspace{0.1cm} & AM \\
        \\
        \hline
        \\
        $\hat{u}_{yz}$    & \hspace{0.1cm} & $k_y^2k_z^2(3k_x^2 - k_y^2)(k_y^2+k_z^2)$ & \hspace{0.1cm} & $^12/^1m$   & \hspace{0.1cm} & FM \\
        \\
        \hline
    \end{tabular}
    }
    \caption{First order correction to the Hamiltonian of the minimal model for the investigated strain cases and corresponding spin point groups (SPGs) of $\mathcal{H}_0 + \mathcal{H}_{\hat{u}}$. The labels for the SPG class are altermagnetic (AM) and ferromagnetic (FM). The three SPGs in the AM class all have $d$-wave symmetry.}
    \label{tab::toy_model_table}
\end{table}

\begin{figure*}
    \centering
    \includegraphics[width=0.9\textwidth]{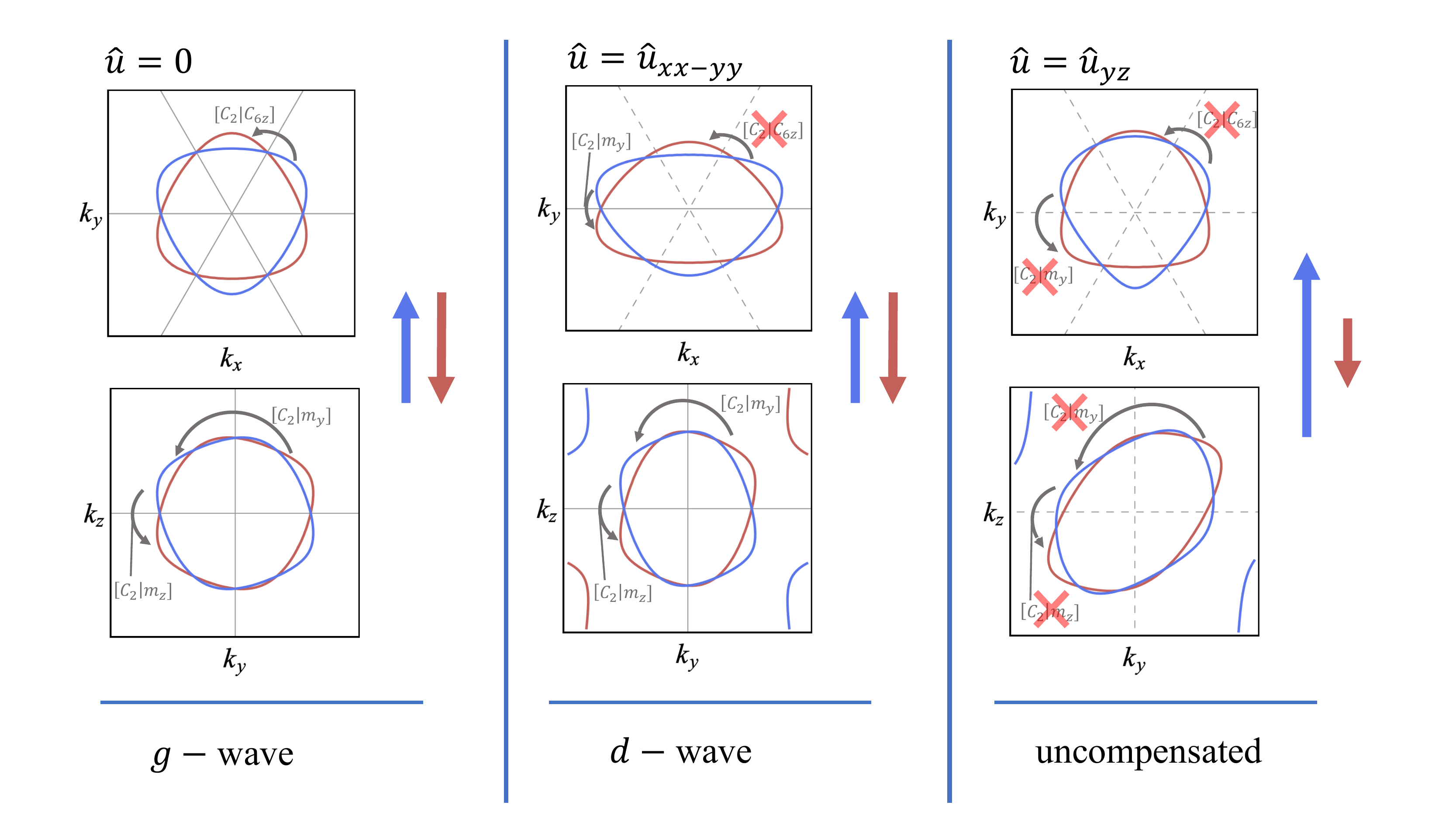}
    \caption{Fermi surface cuts of the hexagonal altermagnetic minimal model when unstrained (left) and under shear strains $\hat{u}_{xx-yy}$ (center) and $\hat{u}_{yz}$ (right). Cuts on the top row show the $k_x,k_y$-plane, with $k_z\neq 0$, and those on the bottom show the $k_y,k_z$-plane, with $k_x\neq 0$. Gray solid lines in the plots indicate the nodal planes, and dashed lines show where nodal planes are broken by strain. Three symmetry operations are labeled, and their preservation or breaking is indicated. Finally the blue and red arrows show the total moment in each spin channel. }
    \label{fig::toy_model_FS}
\end{figure*}

We confirm the results of Table \ref{tab::toy_model_table} with the minimal model. Figure \ref{fig::toy_model_FS} shows cuts of the Fermi surface for the unstrained and strained cases, highlighting the respective $g$-wave, $d$-wave and uncompensated character. For illustrative purposes, the strains have been enhanced to 20\%. The results show for the $\hat{u}_{xx-yy}$ strain how the $g$-wave symmetry in the $k_x,k_y$-plane is broken but a $d$-wave symmetry in the $k_y,k_z$-plane is preserved. On the other hand applying a $\hat{u}_{yz}$ strain removes all nodal surfaces and, nicely visible in the $k_y,k_z$ plane, produces a net magnetic moment, as there is no connection between the spin-up and spin-down channels anymore. We note that for both the transition to the $d$-wave and the uncompensated phase, there is some subtlety to the induced phases, as some of the $g$-wave character remains. This is clearly visible in the Fermi surface cuts of Fig. \ref{fig::toy_model_FS}: the Fermi surfaces still bear traces of the $g$-wave parent structure. For example, although the Fermi surfaces of the middle panels have the $d$-wave symmetry, they look markedly different from the usual pictorial representation of a $d$-wave Fermi surface, which is much simpler. In fact, what we will refer to as $d$-wave here could also be described as $d$-wave + $g$-wave, and similarly, the uncompensated phase as $s$-wave + $g$-wave. 

Nevertheless, these results demonstrate the possibility of controlling the emergence of spin currents and therefore the spin-splitter effect in $g$-wave altermagnets by certain strains that can be determined from the symmetry of the spin elastoconductivity tensor. Further, we have observed that the strained system can lie in an altermagnetic or ferromagnetic spin point group which is determined by the direction of the strain and therefore controllable. To validate and extend these results, we will next examine the strain response of the known altermagnet CrSb through {\em ab initio} calculations.

\section{Ab-initio calculations on CrSb}
\label{sec::CrSb}

CrSb obeys the symmetries of the spin point group $^26/^2m^2m^1m$ \cite{Smejkal2021a}, similar to the model discussed above. Density functional theory calculations reveal that under strain the material exhibits the same spin-symmetry transitions as we saw in that model. To observe the impact of the symmetries and their changes, we consider the band structure of the crystal, both with and without application of strain. 

We have seen in the previous section that certain shear strains reduce the symmetry such that the spin degeneracy at one or more nodal surfaces is lifted. This will lead to extra spin splittings in addition to those already present in the unstrained crystal. Furthermore, these additional splittings are still of altermagnetic origin if two nodal surfaces remain. We start by considering in-plane strains (e.g., $\hat{u}_{xx-yy}$ and $\hat{u}_{xy}$), for which the model predicts that some nodal surfaces are conserved.

\begin{figure*}[th]
    \centering\includegraphics[width=\textwidth]{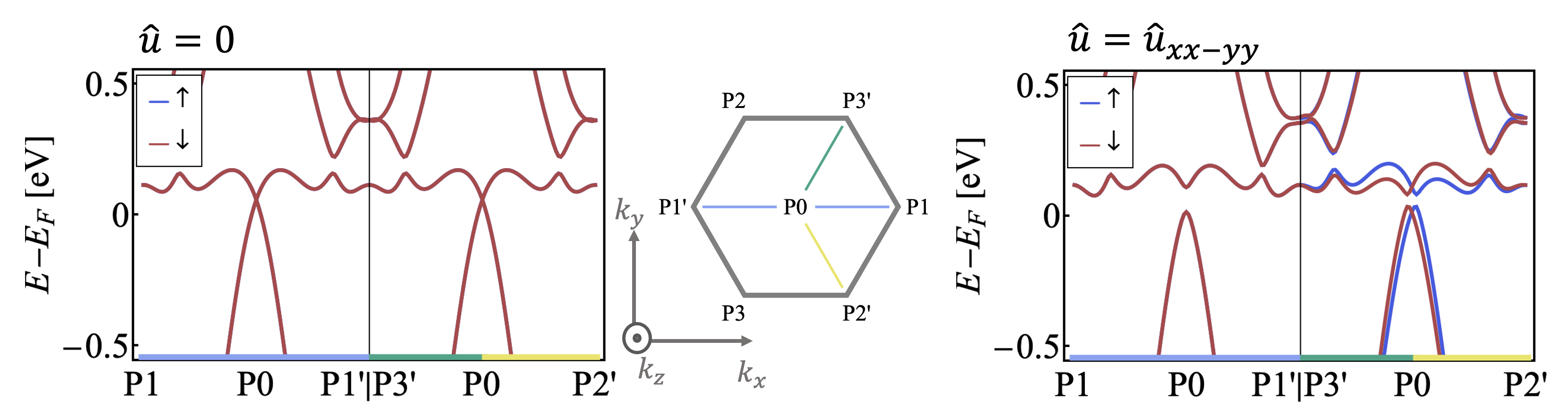}
    \caption{DFT band structure close to the Fermi level, for two paths in the plane $k_z=\pi/2$, for unstrained CrSb (left), and CrSb under 1\% $\hat{u}_{xx-yy}$ shear strain (right). The paths are highlighted in top-view of the Brillouin zone, with the colors corresponding to those on the x-axis of the band-plots.}
    \label{fig::bands_x2-y2}
\end{figure*}

\begin{figure*}[th!]
    \centering\includegraphics[width=\textwidth]{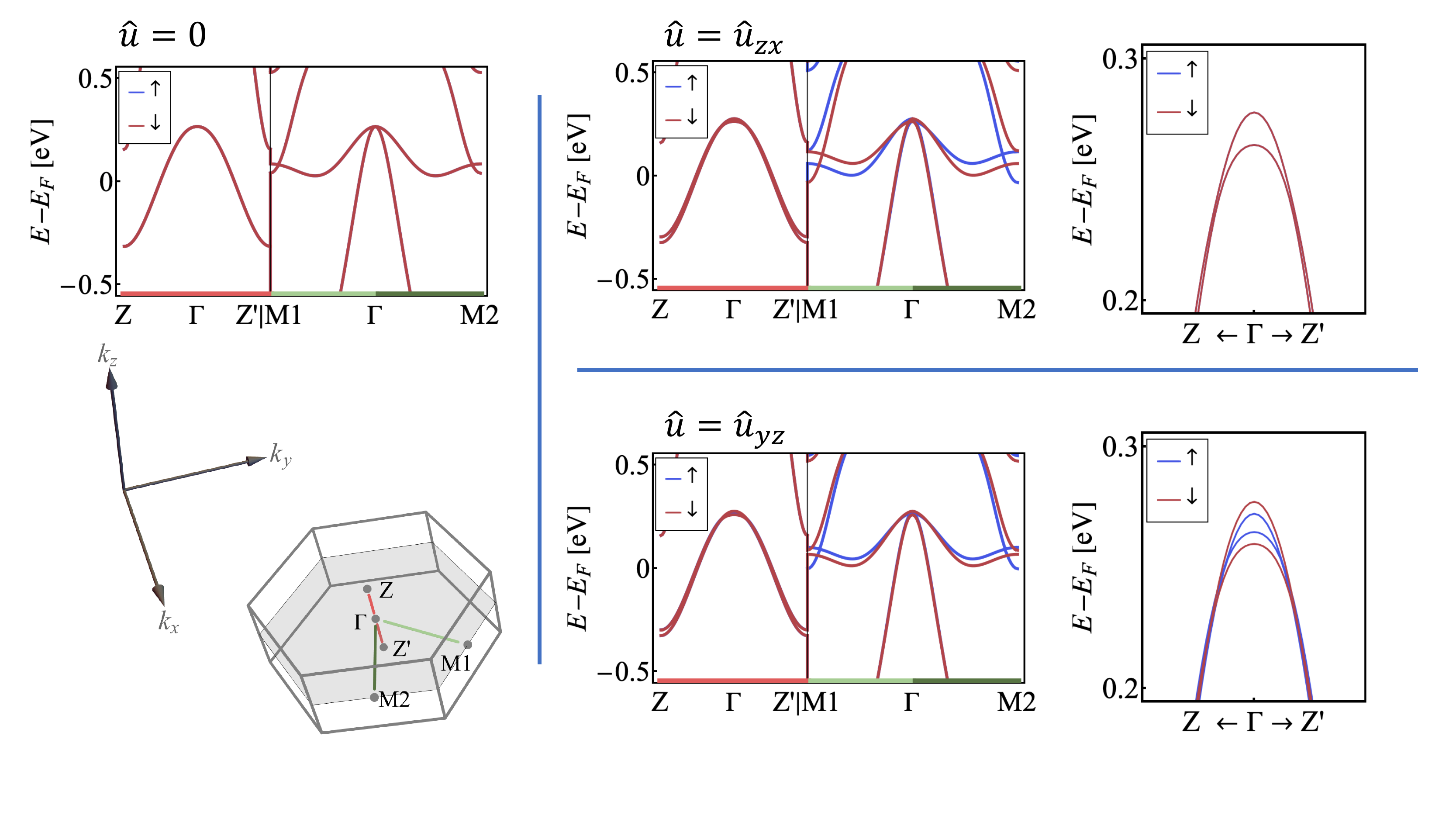}
    \caption{DFT band structure for unstrained (left) and strained ($\hat{u}_{zx}$ and $\hat{u}_{yz}$, middle) CrSb at a strain of 1\%. The paths are highlighted in corresponding colors in the Brillouin zone in the bottom left. For the two strained cases, a detail of the band structure at the $\Gamma$ point is shown on the right.}
    \label{fig::bands_oop}
\end{figure*}

In Fig. \ref{fig::bands_x2-y2}, we show the band structure, obtained with Density Functional Theory (DFT), of unstrained and strained ($\hat{u}_{xx-yy}$) CrSb at 1\% strain, for a path along the hexagonally connected nodal surfaces. To illustrate the effect of the in-plane strain, breaking the in-plane symmetry, we look at two paths parallel to the $z =0$ plane. We set $k_z \neq 0$  because from the symmetry analysis the {$k_x,k_y-$plane} remains a nodal plane. On the left panel of Fig. \ref{fig::bands_x2-y2}, we see that in the unstrained case the bands along P1-P0-P1' and P3'-P0-P2 are identical, since the paths are connected by rotational symmetry. They also show no spin splitting, with the spin-up and spin-down bands lying exactly over each other, as both of these paths are within the nodal planes of the SPG $^26/^2m^2m^1m$. On the other hand, in the right panel we see that at 1\% $\hat{u}_{xx-yy}$ shear strain, the paths are no longer identical, and there is splitting along P3'-P0-P2. This is expected, as under this strain the new SPG is $^2m^2m^1m$ which supports only two nodal surfaces, in this case the $k_x,k_y$- and $k_z,k_x$- plane, and the 3-fold and 6-fold rotational symmetries are broken as well. We see similar behavior for $\hat{u}_{xy}$ strain, though there the SPG under strain is $^22/^2m$ and the nodal surfaces are the $k_xk_y-$plane and a curvilinear surface. We thus confirm the results of the upper part of Table \ref{tab::toy_model_table}, and the prediction of the minimal model: shear strain can induce a sufficient symmetry change in CrSb to create additional spin splittings, consistent with a $d$-wave symmetry. 
Next, we consider the out-of-plane strains.

Figure \ref{fig::bands_oop} shows the bands of unstrained and strained CrSb ($\hat{u}_{zx}$ and $\hat{u}_{yz}$) at 1\% strain for a path along Z-$\Gamma$-Z' and along M1-$\Gamma$-M2, two paths that illustrate the difference in symmetry breaking between the two strains. In the absence of strain the spin is degenerate along both paths. The degeneracy along M1-$\Gamma$-M2 is lifted by both applied strains, but for Z-$\Gamma$-Z' it is different. On the one hand, the nodal surface containing Z-$\Gamma$-Z' is preserved for $\hat{u}_{zx}$, showing similar behavior as $\hat{u}_{xx-yy}$ (Fig. \ref{fig::bands_x2-y2}). On the other hand, applying $\hat{u}_{yz}$ lifts the degeneracy on both nodal surfaces, showing splitting along Z-$\Gamma$-Z' as well. In this case, there are no altermagnetic symmetries anymore, and the material enters a magnetically uncompensated phase. We note that in this uncompensated phase the bands are also spin split at the $\Gamma$-point (see Fig. \ref{fig::bands_oop}), in contrast to any altermagnetic phase, where spin-splitting at $\Gamma$ is forbidden. 
Although this splitting is only of the size of $\sim$10 meV, it switches sign when strain direction is reversed or the spins are flipped, and therefore unlikely to be a feature of numerical inaccuracy. Thus, the band structure shows an uncompensated pattern as predicted. 
{ Band structures for the paths of Fig. \ref{fig::bands_x2-y2} and \ref{fig::bands_oop}, for all four strain cases can be found in the Supplemental Material.}

To better illustrate the breaking or preservation of the different symmetry operations under strain, we also computed the Fermi surfaces of CrSb, by evaluating the spectral function $A^{\uparrow / \downarrow}(\mathbf{k})$ for the two spin channels respectively:

\begin{align}
    A^{\uparrow / \downarrow}(\mathbf{k}) = \frac{1}{\pi}\mathrm{Im}\left\{\mathrm{Tr}\left[\frac{1}{H^{\uparrow / \downarrow}(\mathbf{k})+i\eta}\right]\right\},
\end{align}
for certain cuts in the $k$-space. Here, $H(\mathbf{k}){\uparrow / \downarrow}$ is the Hamiltonian, obtained from a Wannierization procedure of the DFT electronic band structure for the respective spin channel, and $\eta$ is a broadening factor. The spectral function peaks at those k-points where a band crosses the Fermi level, thus allowing us to trace the Fermi surface. In this case, we include all points, where the value of $A(\mathbf{k})$ is larger than a given threshold.

\begin{figure*}[th]
    \centering
   \includegraphics[width=\textwidth]{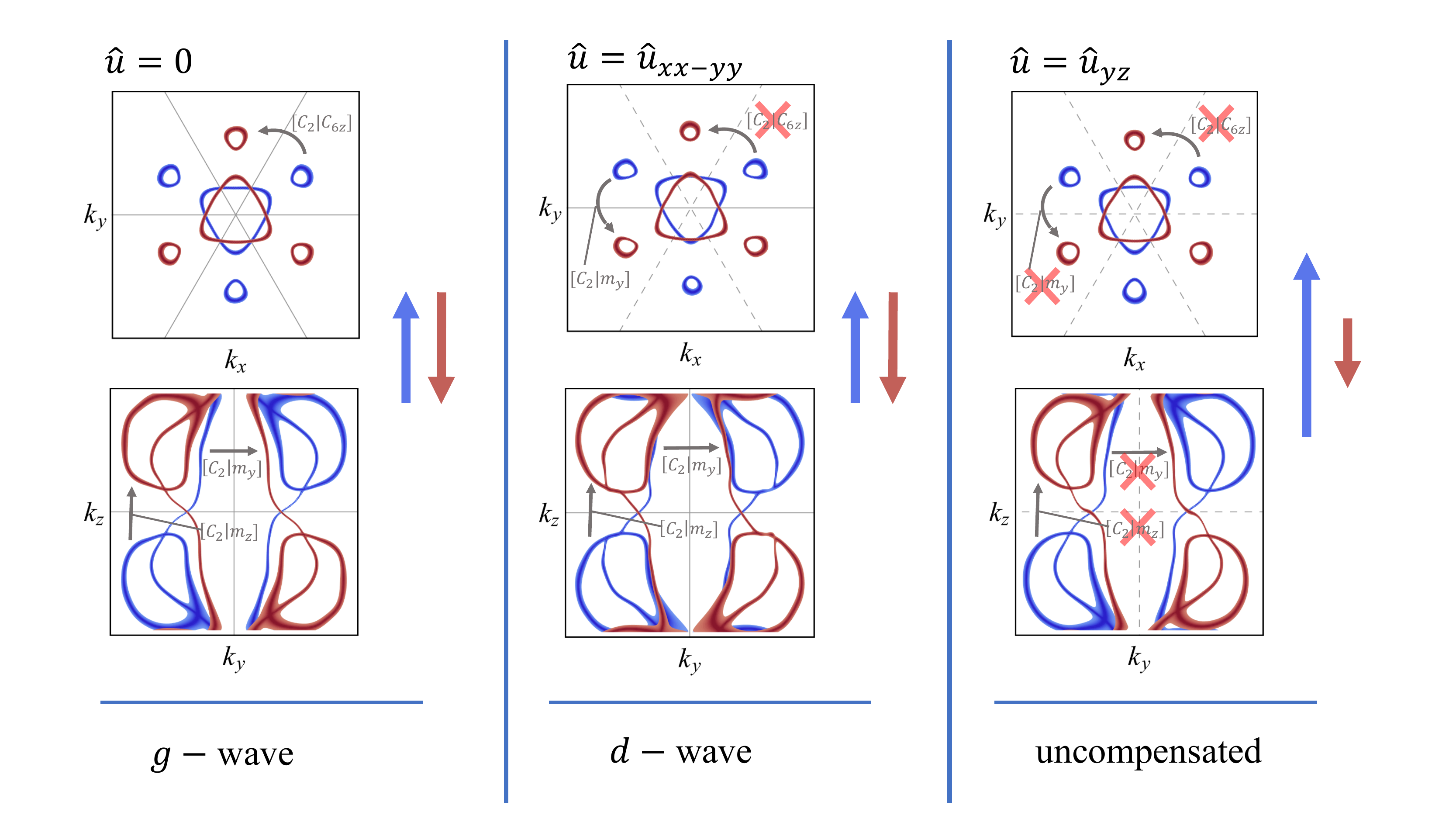}
    \caption{Spin polarized Fermi surface cuts of CrSb when unstrained (left) and under shear strains $\hat{u}_{xx-yy}$ (center) and $\hat{u}_{yz}$ (right) at 1\%. Cuts on the top row show $k_x,k_y$-plane, with $k_z\neq 0$, and those on the bottom show the $k_y,k_z$-plane, with $k_x = 0$. Gray solid lines in the plots indicate the nodal planes, and dashed lines show where nodal planes are broken by strain. Three symmetry operations are labeled, and their preservation or breaking is indicated. Finally the blue and red arrows show the total moment in each spin channel.}
    \label{fig::CrSb_FS}
\end{figure*}

\begin{figure}[th]
    \centering
    \includegraphics[width=.48\textwidth]{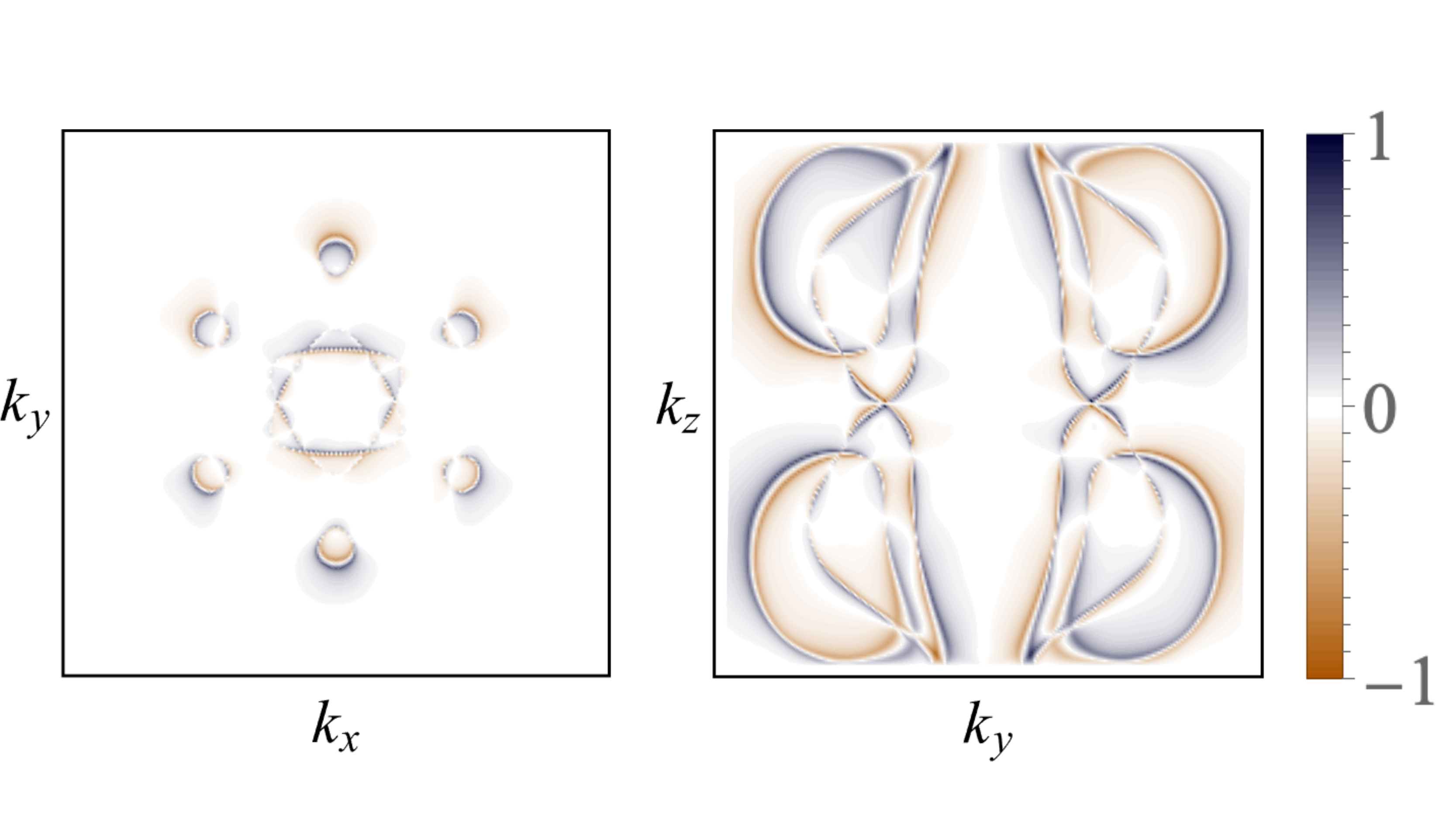}
    \caption{Normalized difference of the spectral functions for out-of-plane strain $\hat{u}_{yz}$ before and after applying $m_y$. Blue color (positive values) corresponds to a band on the Fermi surface before mirroring and orange colors (negative values) correspond to a band after applying the mirror operation. White colors indicate either no or overlapping bands in the Fermi surface at the respective position. }
    \label{fig::CrSb_FS_diff}
\end{figure}

Fig. \ref{fig::CrSb_FS} shows the thus obtained Fermi-surface cuts for CrSb when unstrained, under 1\% $\hat{u}_{xx-yy}$ strain, and under 1\% $\hat{u}_{yz}$ strain. When CrSb is unstrained, we see the $g$-wave symmetry reflected in both Fermi surface cuts. As displayed in the left panel of Fig. \ref{fig::CrSb_FS}, both the expected hexagonal symmetries and the four nodal surfaces characteristic of the $g$-wave splitting are present. Furthermore, the Fermi surfaces of the two magnetic sublattices are connected by $[C_{2}||C_{6z}]$, $[C_2||m_z]$, $[C_2||m_y]$ symmetry. In the middle panels, we see that the $[C_{2}|C_{6z}]$ symmetry is broken under $\hat{u}_{xx-yy}$ strain, but both the $[C_2||m_z]$ and $[C_2||m_y]$ symmetries are preserved. Finally, under $\hat{u}_{yz}$ strain $[C_{2}||C_{6z}]$, $[C_2||m_z]$ and $[C_2||m_y]$ are all broken, as can seen seen on the right panel. Since the effect of broken symmetries on the Fermi surface in the $\hat{u}_{yz}$ case are relativity subtle, we show the difference more clearly in Fig. \ref{fig::CrSb_FS_diff}. Here we display the different between the spectral function $A(\mathbf{k})$, before and after the application of the transposing mirror symmetry $m_y$. Here we see that the Fermi surface is significantly different after the mirror operation, and therefore that this symmetry is broken. So, we see that the Fermi surfaces extracted from the density functional theory calculations reproduce the symmetries of Tab. \ref{tab::toy_model_table} and Fig. \ref{fig::toy_model_FS}, confirming the transition from a $g$-wave altermagnet to either a $d$-wave altermagnet or an uncompensated magnet, depending on the strain conditions.

Up to now, all the discussed results were in the non-relativistic regime, where we did not take SOC into account. However, in the real material, SOC  is of course always present, and we must check that our conclusions hold in this case. In our DFT calculations, we add SOC as a perturbation, and assume that in each of the strain cases the N\'eel vector remains parallel to the $c$-axis (Cartesian $z$-axis). For unstrained CrSb, and each of the different shear strains cases, we calculated the bands and Fermi surfaces. First of all, the bands remain similar, with some moderate change close to the $\Gamma$ point, and the spin splitting pattern does not alter significantly from the non-relativistic case, varying slightly in magnitude only (see Supplementary Material). Additionally, the Fermi surfaces show very similar results as well, displaying the same symmetry-breaking patterns as those without SOC. This indicates that the spin-splitting of CrSb is dominated by the exchange, and that the effect of the strain on the symmetry dominates over any SOC based effects. 

\section{Spin elastoconductivity}
\label{sec::spin-elastoconductivity}
As stated in the introduction, we are interested in how the transport properties of CrSb change under the different strain conditions. Furthermore, for practical applications, it is vital to establish that the emerging effects are of sufficient magnitude. Therefore, we calculate the charge and spin conductivity of strained CrSb.

\begin{figure*}[th]
    \centering
    \includegraphics[width=.492\textwidth]{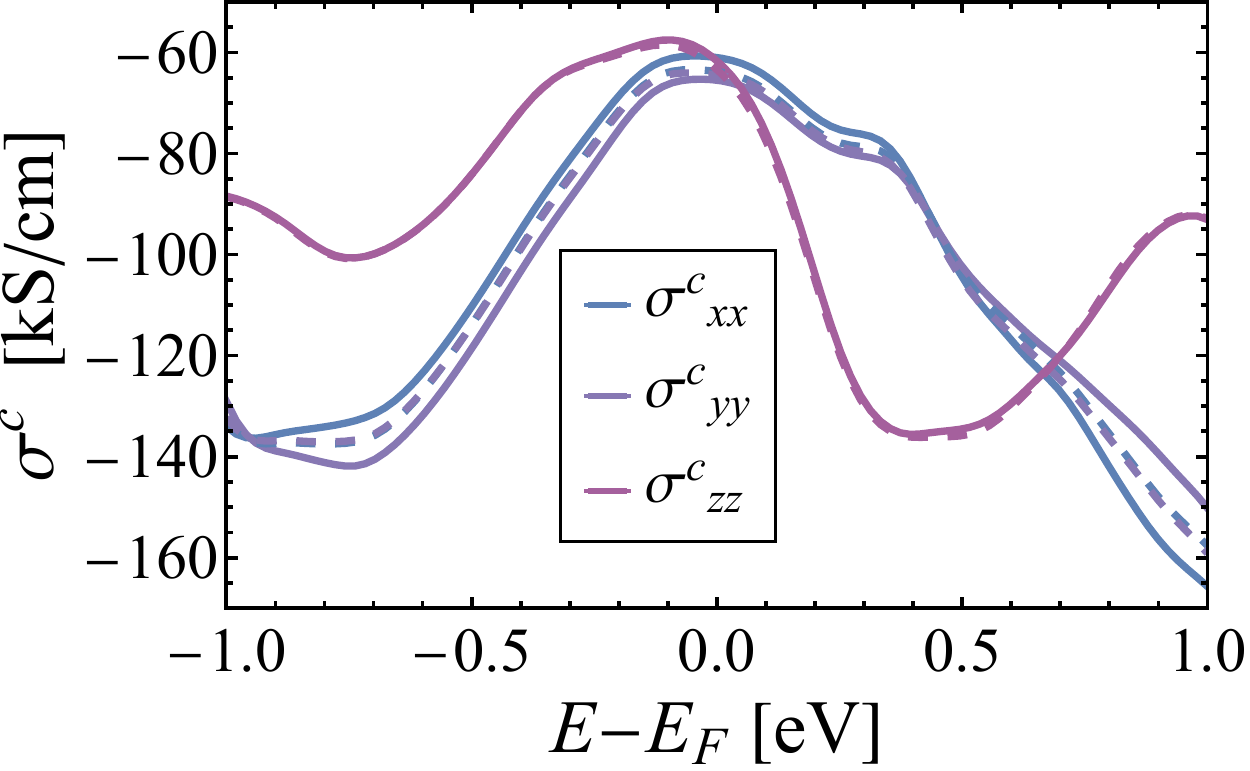}
    \hspace{.025\textwidth}
    \includegraphics[width=.467\textwidth]{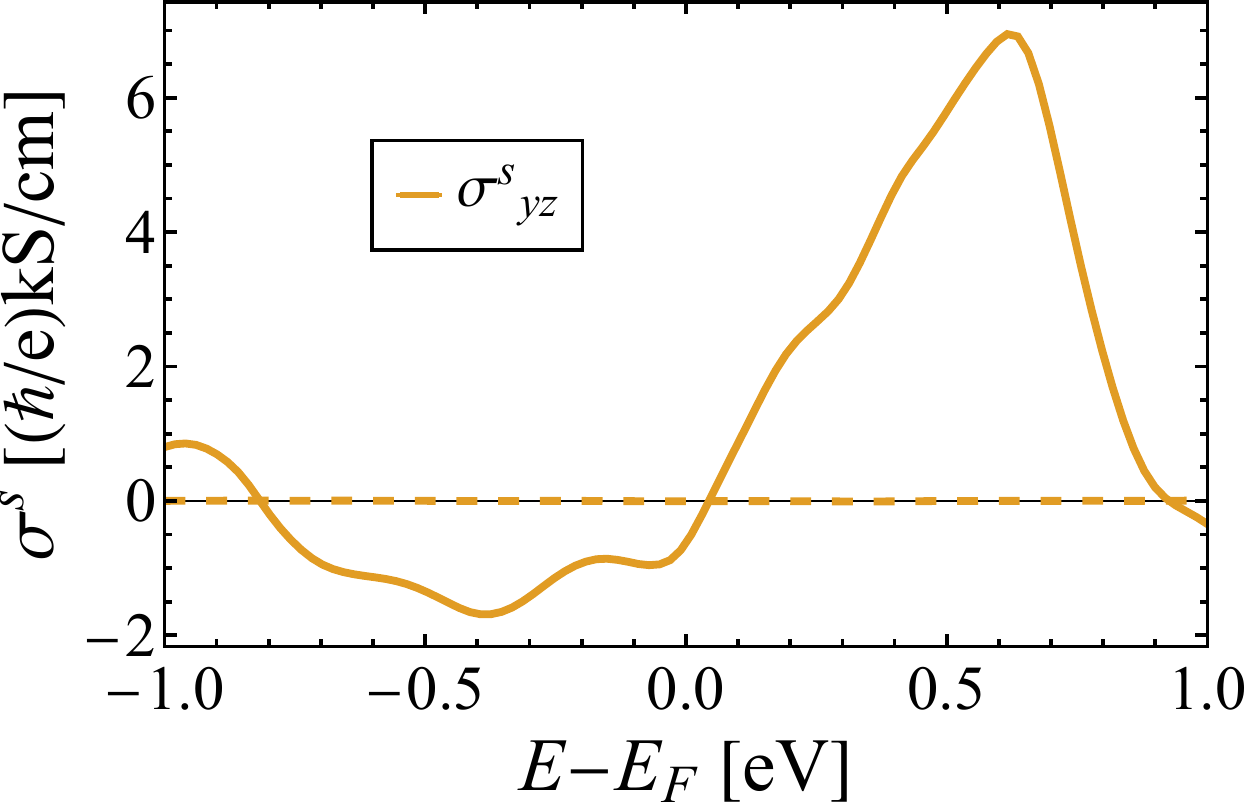}
    \caption{Charge (left) and spin (right) conductivity of strained CrSb with a shear strain ($\hat{u}_{xx-yy}$) of 1\%. dashed lines indicate the values of the unstrained system.}
    \label{fig::currents}
\end{figure*}

In Fig. \ref{fig::currents} we show these for the 1\% $\hat{u}_{xx-yy}$ strain, as an example, and compare them to the unstrained case. We see on the left panel that the charge currents do not change significantly when the strain is applied. However, where the spin conductivity is zero in the unstrained case, a significant contribution arises under the application of shear strain. We note that the emergent current obeys the predicted symmetry and that we obtain similar results for each of the other strain directions.

The spin splitter effect, involves dissipative currents, which are proportional to the relaxation time of the carriers. However, the spin splitter angle (SSA), is independent of this relaxation time (in the clean limit), and it measures the efficiency in the charge-to-spin conversion. The SSA is defined as the angle between the currents of both spin channels when an AM exhibits the spin-splitter effect, i.e., with orthogonal charge and spin currents. For a longitudinal charge current along axis $i$ and orthogonal spin current along axis $j$, the SSA can be defined as:
\begin{align}
    \theta_i^{\,\,\,ij} = \arctan \left( 2 \frac{(2e/\hbar) \sigma^s_{ij}}{\sigma^c_{ii}} \right) . \label{eq:ssa}
\end{align}
where $\sigma^s_{ij}$ is the transverse spin conductivity, and $\sigma^c_{ii}$ is the longitudinal charge conductivity.

\begin{figure}[th]
    \flushleft
    \hspace{-3.9cm}
    \includegraphics[width=.75\textwidth]{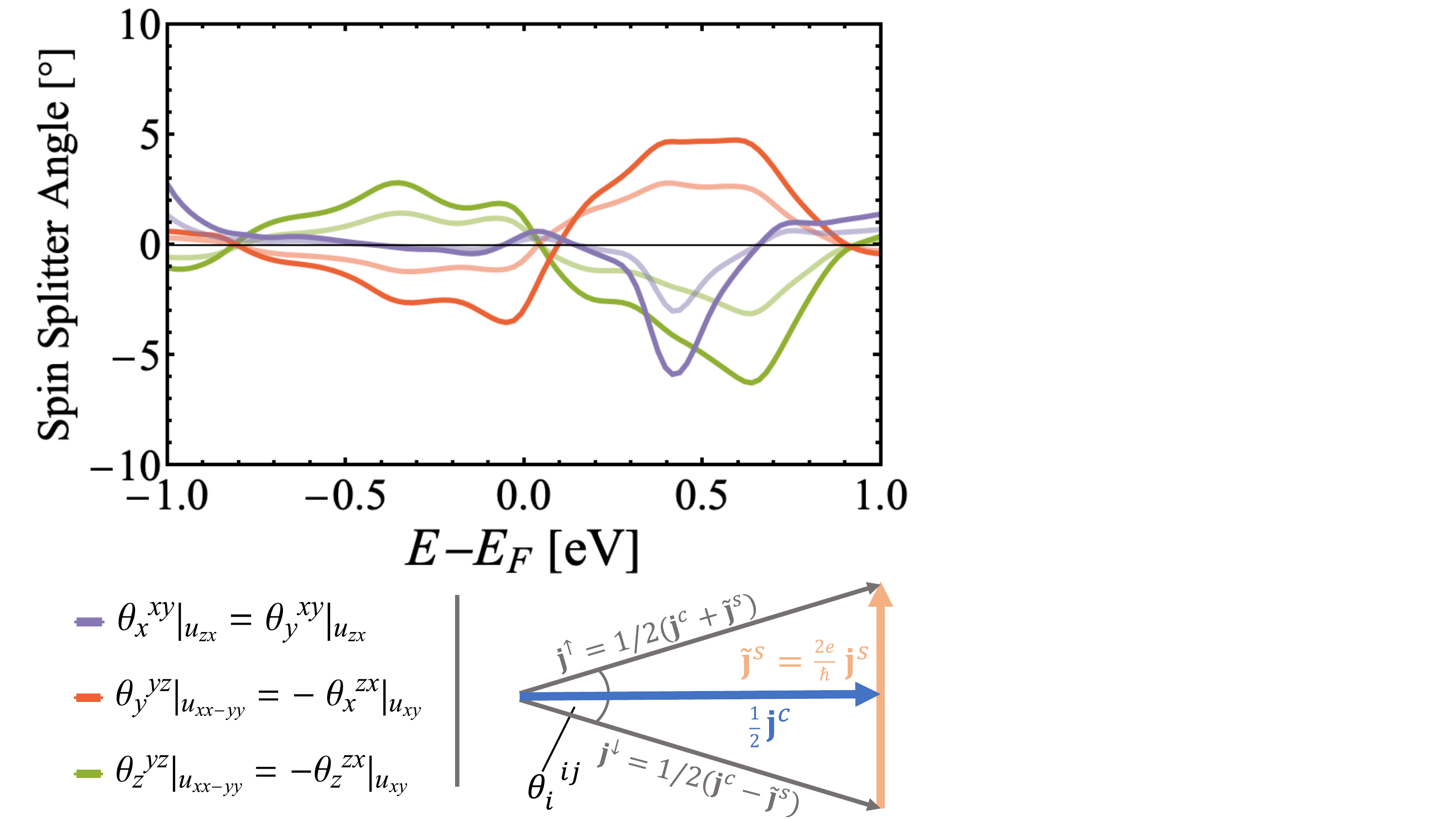}
    \caption{Spin-splitter angle $\theta_{i}^{\,\,\,ij}$ of strained CrSb with 0.5\% (brighter line) and 1\% (darker line) strain.}
    \label{fig::spin_splitter}
\end{figure}

We estimate the SSA for all four strains, and for different charge current directions.
We present the results for the SSA in Fig. \ref{fig::spin_splitter} together with an illustrative reasoning of the formula of Eq. \ref{eq:ssa} to obtain $\theta_i^{\,\,\,ij}$. We see that among the four strains and three current directions, only three independent components arise. This is - again - predicted by the spin elastoconductivity tensor. Furthermore, the SSA reaches up to values of 5\% at 1\% strain from zero spin currents at no strain, which is comfortably in the measurable regime. 
Thus, we see that straining CrSb does not only allow for transitions from a $g$-wave altermagnet to a $d$-wave altermagnet or to an uncompensated magnet, but that, in each of these cases, a non-negligible spin-splitter current emerges, which is forbidden in the unstrained material. These results reveal shear strain as a powerful tool for manipulating the properties of altermagnets.

\section{Conclusion}
\label{sec::conclusion}

In summary, we have established that the $g$-wave altermagnet CrSb can be manipulated with shear strain to undergo a transition to a $d$-wave altermagnet or an uncompensated magnet. Starting from an analysis of the spin symmetries, we have found four strain direction for which such a transition takes place, and, as a consequence, a spin-splitter effect emerges. With a combination of symmetry analysis, a minimal model, and \emph{ab initio} calculations, we have determined which symmetries are broken in each case, and the subsequent consequences for the nodal planes and the spin-splitting. We further have established that these results remain valid in the presence of SOC. Finally, we have shown that with the change in symmetry, spin-splitter currents of observable magnitude emerge. 

In our study, we have explored a wider range of strain directions than considered previously \cite{Belashchenko2024}, aiming for a more complete picture of the effect of shear strain-induced symmetry lowering in $g$-wave altermagnets. We show that strain can serve not only as a means to induce a transition, but also as a precise tool to control the symmetry of the resulting phases, showing three different types of $d$-wave altermagnet and one type of uncompensated magnet for CrSb. Notably, we find that relatively small strain amplitudes are sufficient to produce clear measurable effects. Our findings add to a growing body of both experimental and theoretical research on strain engineering in altermagnetic systems, where different types of strain have been used to stabilize \cite{Reichlova2024, Zhang2025} altermagnetic phases or induce a transition from an antiferromagnetic to altermagnetic phase \cite{Chakraborty2024}, highlighting the broad potential of strain as a tuning parameter. \\
\\
\textit{Acknowledgments} --- We acknowledge funding by 
the Deutsche Forschungsgemeinschaft (DFG, German Research Foundation) -  TRR 288 – 422213477 (project A09 and A12) and TRR 173 – 268565370 (project A03).
 We acknowledge the high-performance computational facility of supercomputer “Mogon” at Johannes Gutenberg-Universität Mainz, Germany. 

\appendix 
\section{Computational Methods}
\label{sec::computational_methods}

Our first-principles calculations based on density functional theory (DFT) were performed in the plane-wave basis as implemented in the Vienna ab-initio simulation package (\textsc{vasp}) \cite{Kresse1996, Kresse1996a} (version 5.4.4), within the collinear local spin generalized gradient approximation (GGA), with the Perdew-Burke-Ernzerhof parametrization of the exchange-correlation \cite{Perdew1996}. The bands of CrSb can be captured well without the inclusion of a Hubbard U \cite{Reimers2024}, so none was applied here. Furthermore SOC was not included, unless stated explicitly. 
The projector-augmented wave pseudopotentials \cite{Blochl1994} (valence electrons: Cr $3p^6 3d^5 4s^1$ without, $3s^2 3p^6 3d^5 4s^1$ with SOC, Sb 2s$^2$2p$^4$, datasets Cr\_pv without, Cr\_sv with SOC, Sb) were used, with a kinetic energy cut-off of 400 eV for the wavefunctions without SOC and 800 eV in the presence of SOC. Brillouin zone integrations were performed using a uniform Monkhorst-Pack $15\times15\times10$ k-point mesh ($15\times15\times12$ with SOC). With these parameters, we obtained a spin moment on the Cr atoms of 2.1 $\mu_B$ (2.2 $\mu_B$ with SOC), slightly reduced with respect to the experimental value of $\sim $2.7 $\mu_B$  \cite{Snow1952, Takei1963}. Starting from an experimental crystal structure \cite{Kallel1974, Reimers2024}, we relaxed the unit cell shape, volume and atomic positions in the absence of SOC and obtained the hexagonal lattice constants $a = 4.073\, \text{\AA}, \,\,\, c = 5.067 \, \text{\AA}$. Here $a$ deviates less than 0.8 \% from the experimental value, while the crystal is somewhat compressed along $c$, which is 7.2\% smaller than the experimental value \cite{Kallel1974, Reimers2024}. Our lattice vectors in terms of Cartesian coordinates are
\begin{align}
    \mathbf{a} = \left( \frac{1}{2}, -\frac{\sqrt{3}a}{2}, 0 \right), \,\,\, \mathbf{b} = \left( \frac{a}{2} , \frac{\sqrt{3}a}{2} , 0 \right), \,\,\, \mathbf{c} = \left(0 , 0 , c \right). \label{eq:vecs}
\end{align} 

In each of the strained cases, we applied the relevant strain matrix to the vectors, and subsequently relaxed only the internal positions of the atoms. We used the exact same atomic positions for the calculations with SOC included, adding it only as a perturbation for the relaxation of the electronic density. 

To obtain the Fermi surfaces and calculate the transport properties we construct the tight-binding model Hamiltonian by using atom-centered Wannier functions within the VASP2WANNIER90 code \cite{Souza2001}. The spin conductivities were calculated with the WannierBerri code package \cite{Tsirkin2021}. For the Fermi surfaces, we use the obtained tight-binding model to evaluate the spectral function using the iterative Green’s function method, as implemented in the WannierTools package \cite{Wu2017b}. 

\bibliography{references}
\end{document}